\documentclass[preprint]{ptephy_v1}

\preprintnumber{RBRC-1247} 

\usepackage{amsmath} 
\usepackage{graphics} 

\usepackage{amssymb,amsmath}
\usepackage[colorlinks=true, pdfstartview=FitV, linkcolor=red, citecolor=blue, urlcolor=blue]{hyperref}
\usepackage[T1]{fontenc}
\usepackage[caption=false]{subfig}

\graphicspath{{./Figures/}}

\newcommand{\diff}{\mathrm{d}}
\newcommand{\p}{\partial}

\newcommand{\Diff}{{\mathcal{D}}}

\newcommand{\be}{\begin{equation}}      
\newcommand{\ee}{\end{equation}}      
\newcommand{\bea}{\begin{eqnarray}}      
\newcommand{\eea}{\end{eqnarray}}

\newcommand{\im}{\mathrm{i}}

\newcommand{\sfU}{\mathsf{U}}
\newcommand{\sfT}{\mathsf{T}}
\newcommand{\calA}{\mathcal{A}}

\newcommand{\calZ}{\mathcal{Z}}

\allowdisplaybreaks[1]
\usepackage[normalem]{ulem}  
\renewcommand\sout{\bgroup \color{red} \ULdepth=-.5ex \ULset}

\begin{document}

\title{
Global inconsistency, 't~Hooft anomaly, and level crossing in quantum mechanics
}


\author[a,b]{Yuta Kikuchi,}
\author[c]{Yuya Tanizaki}

\affil[a]{Department of Physics, Kyoto University, Kyoto 606-8502, Japan}
\affil[b]{Department of Physics and Astronomy, Stony Brook University, Stony Brook, New York 11794-3800, USA}
\affil[c]{RIKEN BNL Research Center, Brookhaven National Laboratory, Upton, New York 11973, USA
\email{yuta.kikuchi@stonybrook.edu, yuya.tanizaki@riken.jp}
}



\begin{abstract}
An 't Hooft anomaly is the obstruction for gauging symmetries, and it constrains possible low-energy behaviors of quantum field theories by excluding trivial infrared theories. 
Global inconsistency condition is recently proposed as a milder condition but is expected to play an almost same role by comparing high symmetry points in the theory space. 
In order to clarify the consequence coming from this new condition, we discuss several quantum mechanical models with topological angles and explicitly compute their energy spectra.
It turns out that the global inconsistency can be saturated not only by the ground-state degeneracy at either of high symmetry points but also by the level crossing (phase transition) separating those high symmetry points.
\end{abstract}

\subjectindex{B31}

\maketitle

\section{Introduction}\label{sec:introduction}

Quantum field theory (QFT) provides a powerful tool unifying the relativity and quantum mechanics in high energy physics as well as the long-range description of many-body physics. Universality of QFT sometimes exhibits common aspects in seemingly quite different systems and allows us to treat them in an analogous fashion~\cite{Wilson:1973jj}. 
Perturbative aspects of QFTs are very well understood, and it gives an approximate result successfully when QFTs are weakly coupled. 
Many examples of our interest are, however, described by strongly-coupled QFTs, and they are very difficult to deal with in general. 
Solution of strongly-coupled QFT is generically unknown unless it is in a special situation, such as in low dimensions, with strong-weak dualities, with certain supersymmetries, etc.  
It is therefore of great importance to give a rigorous statement on QFTs that applies even when QFTs are strongly coupled. 

A key clue is global symmetry of QFT. One cornerstone of traditional many-body physics is Landau's characterization of phases~\cite{landau1937theory,ginzburg1950theory}: different phases realize different symmetries. 
At generic values of coupling constants, the free energy is an analytic function, but some singularities must appear when symmetry breaking pattern changes. 
Another interesting consequence of spontaneous symmetry breaking is the existence of massless bosons called Nambu--Goldstone bosons when a continuous symmetry is spontaneously broken~\cite{Nambu:1960tm, Goldstone:1961eq}. 
As we have seen in these two famous examples, we can give rigorous statements about strongly-coupled  field theories by assuming patterns of spontaneous breaking of global symmetries. 
This is already surprising, but it requires other nonperturbative data to answer the question about whether the symmetry is spontaneously broken or not. 

In certain situations, we can obtain further nonperturbative data even when the system is strongly coupled. Topology related to global symmetries can exclude trivially gapped phase, which is captured by an 't~Hooft anomaly~\cite{tHooft:1979rat, Frishman:1980dq, Coleman:1982yg, Kapustin:2014lwa, Kapustin:2014zva}. 
It would be helpful in the following to discriminate closely related notions called ``anomaly''. 
The anomaly was originally discovered as a pathological feature of QFT, where symmetry of a classical gauge theory is explicitly broken by its quantization~\cite{Adler:1969gk, Bell:1969ts}, or by the path-integral measure~\cite{Fujikawa:1979ay, Fujikawa:1980eg}. 
The famous example is the absence of axial $U(1)$ symmetry in massless quantum electrodynamics (QED) or in massless quantum chromodynamics (QCD) due to the Adler--Bell--Jackiw chiral anomaly. Meanwhile, there is another but related notion called an 't Hooft anomaly. 

%

An 't~Hooft anomaly is defined as an obstruction to promoting the global symmetry to local gauge symmetry~\cite{tHooft:1979rat,Kapustin:2014lwa}. 
We consider a QFT $\mathcal{T}$ with a global symmetry $G$, and let $\calZ[A]$ be the partition function of $\mathcal{T}$ under the background $G$-gauge field $A$. 
We say that $G$ has an 't Hooft anomaly if the partition function $\calZ$ follows the nontrivial transformation law\footnote{Other obstructions to gauging the symmetry exist as shown in Ref.~\cite{Kapustin:2014zva} when the symmetry is discrete, but we do not consider such subtle obstructions in this paper. The anomaly inflow for 't Hooft anomaly not of Dijkgraaf--Witten type is discussed in Ref.~\cite{Thorngren:2015gtw}. }, 
\be
\calZ[A+\diff \theta]=\calZ[A]\exp\left(\im \calA[\theta,A]\right), 
\label{eq:thooft_anomaly}
\ee
under the $G$-gauge transformation $A\mapsto A+\diff \theta$ and $\calA[\theta,A]$ cannot be canceled by local counter terms. 
Especially when $G=G_1\times G_2$, $G_1$ and $G_2$ is said to have a mixed 't~Hooft anomaly if $G_1$ and $G_2$ themselves have no 't~Hooft anomaly but $G_1\times G_2$ has an 't~Hooft anomaly. 
The classic and famous example of an 't~Hooft anomaly is the flavor symmetry $SU(N_f)_L\times SU(N_f)_R\times U(1)_V$ of massless QCD~\cite{tHooft:1979rat}. 
We should emphasize that, although the existence of an 't~Hooft anomaly itself does not mean breaking of symmetries without coupling to background gauge fields, it imposes constraints on low energy dynamics of theories by combining with anomaly matching argument. 
Therefore, 't Hooft anomalies are important nonperturbative data of QFTs.

't Hooft anomaly matching states that the low-energy effective field theory of the QFT  $\mathcal{T}$ must also follow the same transformation law (\ref{eq:thooft_anomaly}) under the background $G$-gauge field $A$ and the $G$-gauge transformation $A\mapsto A+\diff \theta$. 
Original proof of this statement is given, when $G$ is the continuous chiral symmetry, by introducing the spectator chiral fermions canceling the 't Hooft anomaly and by  making the $G$-gauge field $A$ dynamical. Since the coupling of $\mathcal{T}$ to the $G$-gauge field $A$ can be made arbitrarily small, the low-energy effective theory of $\mathcal{T}$ is unaffected by the presence of $A$ and should produce the same phase $\calA[\theta,A]$ under the $G$-gauge transformation in order to cancel the $G$-gauge anomaly from the spectator fermions~\cite{tHooft:1979rat} (See also \cite{Weinberg2,Harvey:2005it} for review). 
Another proof is given by the important observation that the phase functional $\calA[\theta,A]$ can be written as the boundary term of the gauge transformation of a topological $G$-gauge theory in one-higher dimension. 
This is proven when $G$ is the continuous chiral symmetry in even dimension~\cite{Stora:1983ct, Zumino:1983ew}, and it is true in many examples with discrete global symmetries~\cite{Kapustin:2014lwa, Kapustin:2014zva, Cho:2014jfa}. 
When this is true, we can put the theory $\mathcal{T}$ on the boundary manifold of the topological $G$-gauge theory, and then the low-energy effective theory must be able to lie on the same boundary manifold. As a result, the anomaly inflow~\cite{Callan:1984sa} derives the anomaly matching. 
Latest developments on the understanding of topological materials lead discoveries of new 't Hooft anomalies that include discrete syemmetries \cite{Niemi:1983rq,Redlich:1983kn,Redlich:1983dv} or higher-form symmetries \cite{Kapustin:2013uxa,Kapustin:2014gua,Gaiotto:2014kfa} in the context of high energy and condensed matter physics, and they derive nontrivial consequences of low-energy effective theories~\cite{Witten:2015aba,Seiberg:2016rsg,Witten:2016cio,Gaiotto:2017yup,Wang:2017txt, Tanizaki:2017bam, Komargodski:2017dmc, Komargodski:2017smk, Cho:2017fgz,Shimizu:2017asf, Wang:2017loc, Metlitski:2017fmd}.

The question we would like to address in this paper is whether we can derive a nontrivial result when the 't Hooft anomaly is absent. 
In Ref.~\cite{Gaiotto:2017yup}, a new condition, called the global inconsistency of gauging symmetries, is proposed in order to claim the nontrivial consequence similar to the 't Hooft anomaly. 
They considered in that paper about the four dimensional $SU(n)$ Yang~Mills theory at $\theta=\pi$, and the mixed 't Hooft anomaly is found for the center symmetry and $CP$ symmetry when $n$ is even. This derives the spontaneous breaking of $CP$ symmetry at $\theta=\pi$ under a certain assumption (see \cite{Dashen:1970et, Witten:1980sp, tHooft:1981bkw, Ohta:1981ai, Cardy:1981qy, Cardy:1981fd, Wiese:1988qz, Affleck:1991tj, Creutz:1995wf, Creutz:2009kx, Smilga:1998dh, Witten:1998uka, Halperin:1998rc, Mameda:2014cxa} for early related discussions), and the same conclusion is wanted also for the case when $n$ is odd. 
For that purpose, they point out that the local counter terms for gauging the center symmetry at $\theta=0$ and $\theta=\pi$ must be different in order to be compatible with $CP$ symmetry at those points, and this global inconsistency is claimed to lead the same consequence of the 't Hooft anomaly either at $\theta=0$ and $\theta=\pi$: If the phase of one side (say, $\theta=0$) is trivial, then the phase of the other side ($\theta=\pi$) must be nontrivial. 
In Ref.~\cite{Tanizaki:2017bam}, the authors of this paper suggested a new possibility that is compatible with the global inconsistency: The global inconsistency can be satisfied by the phase transition separating those $CP$-symmetric points when the vacua at those points are trivially gapped. The phase structure of the $SU(n)\times SU(n)$ bifundamental gauge theory with finite topological angles is determined under some assumptions with this proposal. 
In this situation, it would be nice to discuss various solvable models with the global inconsistency to check what kinds of possibility can be realized. 

The purpose of this paper is to elucidate significance of global inconsistency as well as mixed 't~Hooft anomaly in rather simple quantum mechanical models.
One of the model is reminiscent of $SU(n)$ Yang~Mills theory, which possesses mixed anomaly for even $n$ and global inconsistency for odd $n$ \cite{Gaiotto:2017yup}.
The similarity was emphasized in Ref.~\cite{Komargodski:2017dmc} in terms of two- and three-dimensional Abelian-Higgs models. 
The other is similar to $SU(n)\times SU(n)$ bifundamental gauge theory, and they also share several properties in common in view of symmetries and anomalies.
We analyze these models in two ways: operator formalism and path integral formalism.
In the former method, we look at the central extension of representations of symmetry groups. In the latter method, we see inconsistency in the local counter term when promoting global symmetries to local gauge redundancies. Although these two methods does not necessarily give the same information about anomalies, we shall see their connection by explicit computation in our models.
Since energy spectrum and corresponding states are calculable, we can clarify consequences of global inconsistency and mixed 't~Hooft anomaly explicitly.

This paper is organized as follows: In Section \ref{sec:global_inconsistency}, we explain a general idea of the global inconsistency in comparison with the 't~Hooft anomaly.
In Section \ref{sec:model1}, we discuss a particle moving around a circle with a periodic potential. We see how to detect mixed anomaly and global inconsistency in the system and discuss consequences on the energy spectrum. 
In Section \ref{sec:model2}, we add another variable to the model discussed in Section \ref{sec:model1} to mimic the $SU(n)\times SU(n)$ bifundamental gauge theory at finite $\theta$ angles. The global inconsistency plays an even more important role in this model  and we present the resultant energy spectrum and its interpretations. 
We give conclusions in Section \ref{sec:conclusion}.

\section{Global inconsistency of gauging symmetries}\label{sec:global_inconsistency}

In this section, we define the global inconsistency condition, proposed in Ref.~\cite{Gaiotto:2017yup}, in the general context of QFT. 

We consider a QFT $\mathcal{T}$ parametrized by continuous parameters $\vec{g}=(g^1,g^2,\cdots)$ such as mass parameters, coupling constants, theta angles, and so on, which is described by a partition function ${\calZ}_{\vec{g}}$.
At generic values of $\vec{g}$, the QFT $\mathcal{T}(\vec{g})$ has the global symmetry $G$, and we assume that $G$ has no 't Hooft anomaly. 
By this assumption, we can couple the theory $\mathcal{T}(\vec{g})$ to the background $G$-gauge field without breaking the invariance under the $G$-gauge transformation. 
In this process, the topological $G$-gauge theory on the same dimension is introduced, and the parameter space is extended by new couplings $\vec{k}$ of the topological $G$-gauge theory. 
Some of them might be continuous but the other of them will be quantized to ensure the $G$-gauge invariance, and we assume, for simplicity, that all the new couplings $\vec{k}$ is quantized to discrete values\footnote{An example of the discrete parameter $\vec{k}$ is the level of the Chern-Simons theory. }. 
We denote the partition function under the background $G$-gauge field $A$ as $\calZ_{\vec{g},\vec{k}}[A]$, and it satisfies
\be
\calZ_{\vec{g},\vec{k}}[A+\diff \theta]=\calZ_{\vec{g},\vec{k}}[A]
\ee
under the $G$-gauge transformation $A\mapsto A+\diff \theta$. When making the $G$-gauge field $A$ dynamical, we call the obtained theory as $(\mathcal{T}(\vec{g})/G)_{\vec{k}}$, and the global symmetry disappears at generic point of $\vec{g}$. 


Although the symmetry of the theory $\mathcal{T}(\vec{g})$ is $G$ for generic $\vec{g}$, it may be enhanced to other group at special points. 
Let $\vec{g}_1$ and $\vec{g}_2$ be such special points, where the symmetry is enhanced to $G\times H$ by the group $H$, and we shall call these points $\vec{g}_1$ and $\vec{g}_2$ as high symmetry points. 
We restrict our attention to the case where $G\times H$ has no 't Hooft anomaly both at $\vec{g}_1$ and $\vec{g}_2$. 
In this setting, the global inconsistency condition is defined as follows: There exists no $\vec{k}$ such that $\calZ_{\vec{g},\vec{k}}[A]$ is compatible with the $H$-gauge invariance both at $\vec{g}_1$ and $\vec{g}_2$. 

We shall take a closer look at the global inconsistency condition. 
Since there is no 't Hooft anomaly of $G\times H$ at $\vec{g}_i$ ($i=1,2$), there exists $\vec{k}_i$ such that  
\be
\calZ_{\vec{g}_i,\vec{k}_i}[h\cdot A]=\calZ_{\vec{g}_i,\vec{k}_i}[A], 
\ee
where $h\cdot A$ is the transformation of $G$-gauge field $A$ by $h\in H$. 
The condition for the global inconsistency states that $\vec{k}_1\not=\vec{k}_2$. 
When $\vec{k}=\vec{k}_1$ is chosen, the symmetry $H$ at $\vec{g}_2$ is explicitly broken as 
\be
\calZ_{\vec{g}_2,\vec{k}_1}[h\cdot A]=\calZ_{\vec{g}_2,\vec{k}_1}[A]\exp\left(\im \calA_{\vec{g}_2,\vec{k}_1}[h,A]\right) 
\ee
for some phase functional $\calA_{\vec{g}_2,\vec{k}_1}$. 
Therefore, $(\mathcal{T}(\vec{g}_1)/G)_{\vec{k}_1}$ has the symmetry $H$, but $(\mathcal{T}(\vec{g})/G)_{\vec{k}_1}$ has no symmetry including $\vec{g}=\vec{g}_2$. 
The similar equation, 
\be
\calZ_{\vec{g}_1,\vec{k}_2}[h\cdot A]=\calZ_{\vec{g}_1,\vec{k}_2}[A]\exp\left(\im \calA_{\vec{g}_1,\vec{k}_2}[h,A]\right), 
\ee
is true at $\vec{g}_1$ when $\vec{k}=\vec{k}_2$ is chosen: 
\begin{figure}[t]
\centering
\includegraphics[scale=.6]{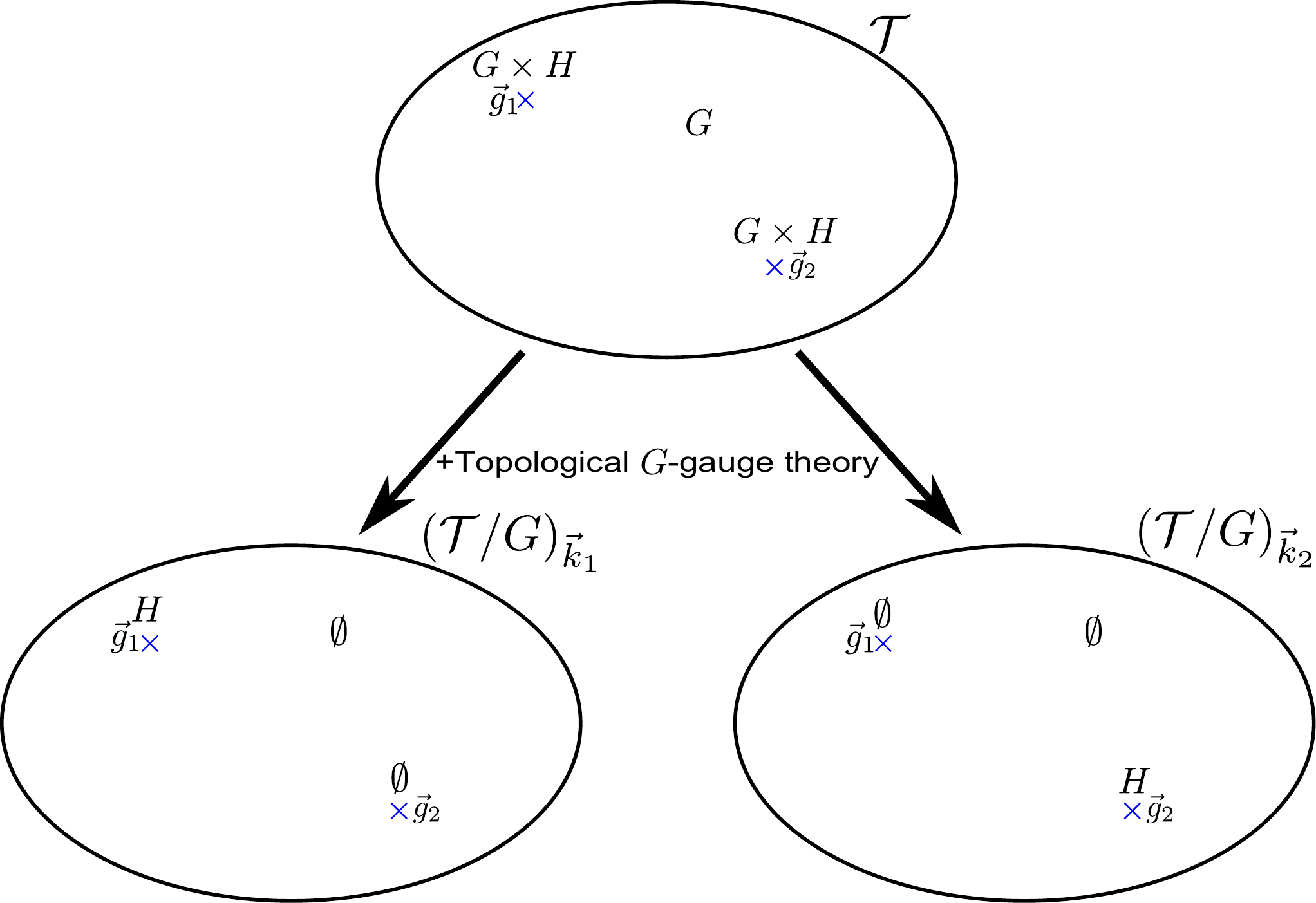}
\caption{The schematic figure illustrating the global inconsistency in the space of coupling constants $\vec{g}$. In the original theory $\mathcal{T}$, symmetry $G$ exists at generic couplings $\vec{g}$ and it is enhanced by $H$ at $\vec{g}_1$ and $\vec{g}_2$. To gauge the symmetry $G$, $\mathcal{T}$ is coupled to the topological $G$-gauge theory with the discrete parameter $\vec{k}$. 
In $(\mathcal{T}/G)_{\vec{k}_1}$, the symmetry is absent except at $\vec{g}=\vec{g}_1$. In $(\mathcal{T}/G)_{\vec{k}_2}$, the symmetry is absent except at $\vec{g}=\vec{g}_2$. }
\label{fig:global}
\end{figure}
$(\mathcal{T}(\vec{g}_2)/G)_{\vec{k}_2}$ has the symmetry $H$, but $(\mathcal{T}(\vec{g})/G)_{\vec{k}_2}$ has no symmetry including $\vec{g}=\vec{g}_1$. 
It should be noted that $\vec{k}$ cannot be chosen individually at each point because the two points are continuously connected in the parameter space and $\vec{k}$, being a discrete parameter, does not change discontinuously on the path connecting the points\footnote{If $\vec{k}$ contains continuous parameters, the corresponding condition is replaced as follows: The global inconsistency exists if there is no connected component of the $\vec{k}$ space that respects full symmetries at both $\vec{g}_1$ and $\vec{g}_2$. }.
This situation is schematically shown in Fig.~\ref{fig:global}.

It should be emphasized that the inconsistent points have to be connected continuously in parameter space.
When there is the global inconsistency between $\vec{g}_1$ and $\vec{g}_2$, we claim that 
\begin{itemize}
\item The vacuum either of $\mathcal{T}(\vec{g}_1)$ or of $\mathcal{T}(\vec{g}_2)$ is nontrivial\footnote{The vacuum is called nontrivial in this paper if the theory is either gappless, with spontaneous symmetry breaking, or with topological order. }, or 
\item $\vec{g}_1$ and $\vec{g}_2$ are separated by the phase transition. 
\end{itemize} 
When the first statement is realized, the global inconsistency condition shows the existence of the nontrivial phase at one of the high symmetry points. 
Meanwhile, the second statement suggests that the global inconsistency is automatically satisfied if there is a phase transition separating the high symmetry points where the discrete parameter $k$ may jump.
This aspect makes the global inconsistency a milder obstruction than the 't~Hooft anomaly and an important corollary is that the existence of global inconsistency does not necessarily lead to nontrivial infrared theory at high symmetry points.

We shall see how these arguments work explicitly by looking at several quantum mechanical examples in the rest of this paper. In all the quantum mechanical examples with the global inconsistency, one of the above conclusions is realized. Moreover, we will find that both possibilities are realized in certain quantum mechanical models.

\section{Quantum mechanics of a particle on $S^1$}\label{sec:model1}

We consider the quantum mechanics on a circle $S^1=\mathbb{R}/2\pi \mathbb{Z}$ with the topological $\theta$ term, describing a particle with unit mass moving on a ring of unit radius. $\theta$ term arises due to the flux threading the ring. 
The Euclidean classical action is 
\be
\label{eq:model}
S[q]=\int\diff \tau \left[{1\over 2}\dot{q}^2+V(n q)\right]-{\im\theta\over 2\pi}\int\diff q.
\ee
The potential $V(x)$ is an arbitrary $2\pi$ periodic smooth function, $V(x+2\pi)=V(x)$, and it can be represented as the Fourier series,
\be
V(x)=\sum_{\ell \ge 1}\lambda_\ell \cos(\ell x+\alpha_\ell). 
\label{eq:potential1}
\ee
Each $q$ is the map $q:S^1_{\beta}\to \mathbb{R}/2\pi\mathbb{Z}$, where $S^1_{\beta}$ is the circle with the circumference $\beta$, $\dot{q}_i={\diff q_i/\diff \tau}$, and $n\ge 2$ is an integer. 
The set of parameters is $\vec{g}=(\theta,\lambda_1,\ldots, \alpha_1,\ldots)$, and we often denote only $\theta$ instead of $\vec{g}$ since the most important parameter in our discussion is $\theta$. 
The parameter $\theta$ is identified with $\theta+2\pi$ because $\int\diff q\in2\pi\mathbb{Z}$. The partition function $\calZ_{\theta}$ is defined by the path integral,
\be
\calZ_{\theta}=\int \Diff q \exp\left(-S[q]\right). 
\ee
In the operator formalism, the Hamiltonian of this system is given by 
\be
\hat{H}(\hat{p},\hat{q})={1\over 2}\left(\hat{p}-{\theta\over 2\pi}\right)^2+V(n\hat{q}), 
\ee
where $[\hat{q},\hat{p}]=\im$ and the Hilbert space $\mathcal{H}$ is the set of  $2\pi$-periodic $L^2$-functions; the partition function is $\calZ_{\theta}=\mathrm{tr}_{\mathcal{H}}[\exp(-\beta\hat{H})]$.

The goal of this section is to figure out the consequences of mixed 't~Hooft anomaly and global inconsistency in this model.
The aspect of the 't Hooft anomaly for this model is already discussed in detail when $n=2$ and $\alpha_\ell=0$ in Appendix of Ref.~\cite{Gaiotto:2017yup}, so this section partly contains the review of known results. 
Still, we would like to start with this model since it is the simplest case where the global inconsistency shows up when $n$ is odd. 
We will indeed see that (non-accidental) level crossings appearing in the energy spectrum can be explained in terms not only of 't~Hooft anomalies but also of global inconsistency.

\subsection{Symmetries, central extension, and global inconsistency}
\label{sec:model1_symmetry}

The system (\ref{eq:model}) has the $\mathbb{Z}_n$ symmetry, generated by 
\be
 \label{eq:Zn_transf}
 \sfU : q(\tau)\mapsto q(\tau)+\frac{2\pi }{n}.
\ee
Since $q(\tau)$ and $q(\tau)+2\pi$ is identified on the circle, $\sfU^n=1$. Quantum mechanically, the symmetry operator $\sfU$ can be realized as 
\be
 \label{eq:Zn_operator}
 \sfU=\exp\left(\im{2\pi \over n}\hat{p}\right), 
\ee
and it is easy to check that $\sfU \hat{H} \sfU^{-1}=\hat{H}$ for any $\theta$. 
We take this convention for $\sfU$ in the following. 

The symmetry of the system is $\mathbb{Z}_n$ for generic $\theta$, but there are additional symmetry at $\theta=0,\pi$. These two points are the high symmetry points of (\ref{eq:model}), and we have the time reversal symmetry $\sfT$, 
\be
\sfT: q(\tau)\mapsto q(-\tau),\; \dot{q}(\tau)\mapsto -\dot{q}(-\tau).
\label{eq:time_reversal_q}
\ee
At $\theta=0$, the action $S$ is quadratic in $\dot{q}$, and this symmetry exists trivially. At generic $\theta$, the topological term is linear in $\dot{q}$ and the time reversal symmetry is absent. 
At $\theta=\pi$, if we perform this transformation, the topological term changes as 
\be
-{\im \over 2}\int \diff q\mapsto {\im\over 2}\int \diff q=-{\im\over 2}\int \diff q+\im \int \diff q.
\ee
Since $\int \diff q \in 2\pi \mathbb{Z}$, the path-integral weight $\exp(-S)$ does not change under $\sfT$. Therefore, the time reversal is the symmetry also at $\theta=\pi$. 
%

Let us study the commutation relation of $\sfU$ and $\sfT$~\cite{Gaiotto:2017yup}. 
Two important conditions, $\sfT \hat{H} \sfT^{-1}=\hat{H}$ and $\sfT \im \sfT^{-1}=-\im$, can be satisfied by  
\bea
 \sfT \hat{q} \sfT^{-1}=\hat{q},
 \hspace{3mm} \sfT \hat{p} \sfT^{-1}=\left\{
 \begin{array}{ll}
  -\hat{p} &(\theta=0),
  \\
  -\hat{p}+1 \ \  &(\theta=\pi).
 \end{array}
 \right.
\eea
If we choose the coordinate basis (i.e. $\hat{q}=q$ and $\hat{p}=-\im \p_q$), we can realize $\sfT$ as $\sfT=\mathcal{K}$ at $\theta=0$, and $\sfT=\exp(\im q)\mathcal{K}$ at $\theta=\pi$, where $\mathcal{K}$ is the complex conjugation. 
%
Using the expression~(\ref{eq:Zn_operator}) and the above commutation relation for $\sfT$, we find that 
\bea
\label{eq:com_rel_1}
\sfT\sfU\sfT^{-1}=\left\{
 \begin{array}{ll}
  \sfU, &(\theta=0),
  \\
  \mathrm{e}^{-{2\pi\im/n}}\sfU, \ \  &(\theta=\pi).
 \end{array}
 \right.
\eea
We have several remarks on the central extension of symmetry group based on the commutation relations \eqref{eq:com_rel_1}.
At $\theta=0$, the $\mathbb{Z}_n$ transformation and time reversal ($\mathbb{Z}_2$) transformation commute as we expected from the enhanced symmetry $\mathbb{Z}_n\times\mathbb{Z}_2$.
However, at $\theta=\pi$ we have an additional phase factor, which may or may not be absorbed by redefining the operator properly. The symmetry group $\mathbb{Z}_n\times\mathbb{Z}_2$ is said to be centrally extended when there is no proper redefinition to absorb the phase factor, which is the central element. Let us redefine the operator by $\sfU'\equiv\mathrm{e}^{-\frac{2\pi\im k}{n}}\sfU$ for some integer $k$. Substituting $\sfU'$ back into the second commutation relation \eqref{eq:com_rel_1}, we obtain
\be
 \sfT \sfU' \sfT^{-1} = \exp\left(\frac{2\pi\im}{n}(2k-1)\right)\sfU'.
\ee
Hence the phase factor can be absorbed when the following condition is satisfied:
\be
\label{eq:condition_k}
k=-k+1\ \ (\mathrm{mod}\ n).
\ee 
Since there is no solution for $k$ to be an integer when $n\in2\mathbb{Z}$, the symmetry group is centrally extended. If we try to redefine the operator with a solution of \eqref{eq:condition_k}, which is a half integer for even $n$, the redefined operator $\sfU'$ satisfies ${\sfU'}^n=-1$ unlike ${\sfU}^n=1$. 
This means that we get a double cover of the original symmetry group $\mathbb{Z}_n\times\mathbb{Z}_2$. We shall see in the next section that this is the consequence of the mixed 't~Hooft anomaly between $\mathbb{Z}_n$ and the time-reversal symmetry~\cite{Gaiotto:2017yup}. When $n\in2\mathbb{Z}+1$, we can redefine the operator $\sfU'$ by choosing $k=(n+1)/2$, which is an integer. Since we succeeded in defining $\sfU'$ with maintaining ${\sfU'}^n=1$, there is no central extension for odd $n$.
This is not the end of story. Although there is no central extension at $\theta=0$ and $\pi$ separately for odd integer $n$, we cannot avoid the central extension at $\theta=0$ and $\pi$ simultaneously by choosing a common operator $\sfU$ (or $\sfU'$). This fact implies the global inconsistency. 

Let us discuss how the above argument constraints the energy spectrum. First let us consider the case when $n\in 2\mathbb{Z}$. 
Let us ask whether there exists a simultaneous eigenstate of $\sfU$ and $\sfT$ at $\theta=\pi$. 
We assume for contradiction that such a state exists and denote it by $|\psi\rangle$. By assumption, we can set 
\be
\label{eq:states}
\sfU |\psi\rangle=\mathrm{e}^{{2\pi \im}k/n}|\psi\rangle,\ \ 
\sfT |\psi\rangle=\eta|\psi\rangle. 
\ee
Here $k\in\mathbb{Z}$, because $\sfU^n=1$ on $\mathcal{H}$. 
Using the commutation relation (\ref{eq:com_rel_1}), we obtain 
\be
\exp\left({2\pi \im\over n}k\right)=\exp\left({2\pi \im\over n}(1-k)\right).
\label{eq:condition_consistency}
\ee
This can be rewritten as \eqref{eq:condition_k}.
When $n$ is even, this does not have any integer solutions: The simultaneous eigenstate of $\sfU$ and $\sfT$ cannot exist at $\theta=\pi$, and all the energy eigenvalues is two-fold degenerate. 

Next, let us consider the case when $n\in 2\mathbb{Z}+1$. In this case, we shall find no 't Hooft anomaly, and thus the simultaneous eigenstate can exist at $\theta=\pi$. Indeed, we obtain the same condition \eqref{eq:condition_consistency} for the simultaneous eigenstates of $\sfU$ and $\sfT$ at $\theta=\pi$, and the possible $\mathbb{Z}_n$ charge is determined as $k=(n+1)/2$ modulo $n$ when $n$ is an odd integer. 
Even in this situation, the global inconsistency between $\theta=0$ and $\pi$ can derive a nontrivial result: No states can be singlet both at $\theta=0$ and $\theta=\pi$. 
Let $|\psi_0\rangle$ be a simultaneous eigenstate of $\sfU$ and $\sfT$ at $\theta=0$, then the similar computation shows that 
\be
\sfU |\psi_0\rangle = |\psi_0\rangle. 
\ee
Let $|\psi_\pi\rangle$ be a simultaneous eigenstate of $\sfU$ and $\sfT$ at $\theta=\pi$, then the above argument has shown that 
\be
\sfU |\psi_\pi\rangle = \exp\left({2\pi \im\over n}{n+1\over 2}\right)|\psi_\pi\rangle. 
\ee
Since $|\psi_0\rangle$ and $|\psi_\pi\rangle$ have different $\mathbb{Z}_n$ charge, those states cannot be continuously connected by changing the parameter $\theta$ of the theory. 
In other words, the $\sfT$-invariant states at $\theta=0$ break $\sfT$ at $\theta=\pi$, and vice versa. 

\begin{figure}[t]
\centering
\begin{minipage}{.49\textwidth}
\subfloat[$n=4$]{
\includegraphics[scale=0.8]{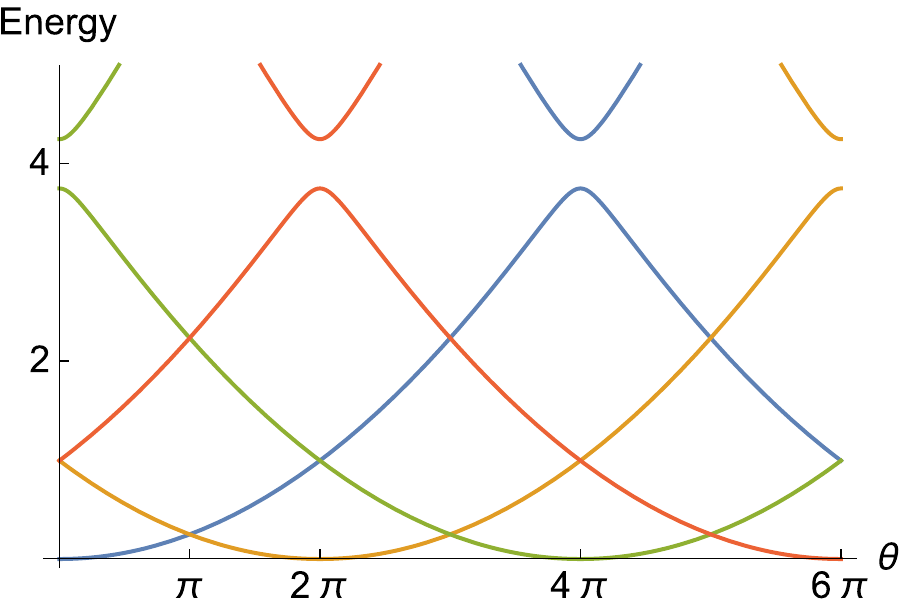}
\label{fig:Z4}
}\end{minipage}\
\begin{minipage}{.49\textwidth}
\subfloat[$n=3$]{
\includegraphics[scale=0.8]{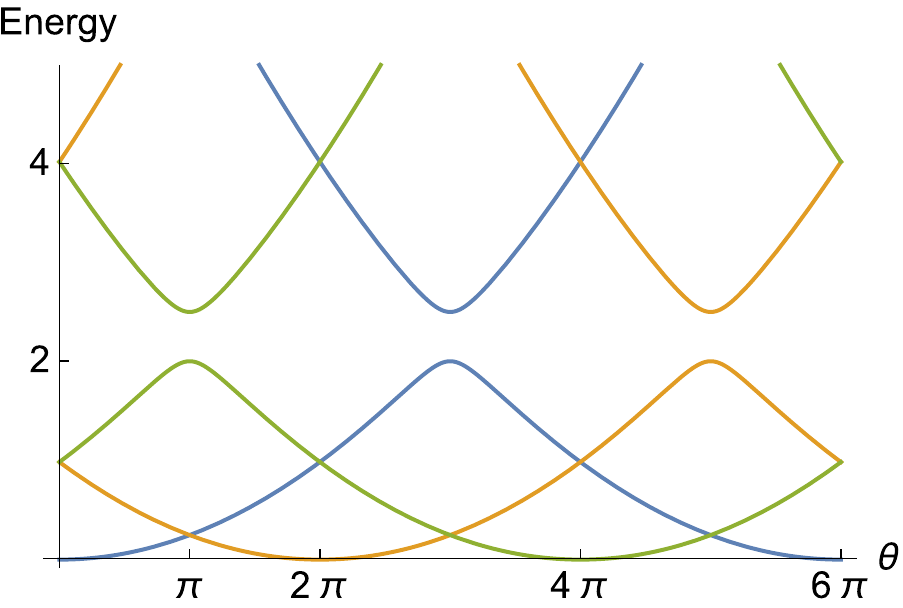}
\label{fig:Z3}
}\end{minipage}
\caption{Energy levels as functions of $\theta$ with $\lambda=0.5$ in \eqref{eq:cos_int} for $\mathbb{Z}_4$ and $\mathbb{Z}_3$ symmetric cases, respectively. Each color corresponds to different $\mathbb{Z}_n$ charge. 
(a) Every state forms a pair at $\theta=\pi, 3\pi, 5\pi$, which is a consequence of the t'~Hooft anomaly. 
(b) Not every state forms a pair at $\theta=\pi, 3\pi, 5\pi$. But, a singlet state at $\theta=0$ are not continuously connected to a singlet state at $\theta=\pi$, which is a consequence of the global inconsistency.}
\label{fig:Zn}
\end{figure}

To make the above arguments more convincing, let us compute the energy spectrum explicitly for the potential, 
\be
\label{eq:cos_int}
V(nq)=\lambda \cos(n q). 
\ee
Figure~\ref{fig:Zn} shows the energy spectra for the cases $n=4$ and $n=3$ that are computed numerically by diagonalizing the Hamiltonian. 

As we can see in Fig.~\ref{fig:Z4}, no sates can be singlet at $\theta=\pi$ when $n=4$ and this is expected  because of the nontrivial commutation relation between $\sfU$ and $\sfT$. 
When $n=3$, there are singlet states at $\theta=\pi$ as shown in Fig.~\ref{fig:Z3}, and this is allowed from the commutation relation. 
The point is that a singlet state at $\theta=0$ and the singlet state at $\theta=\pi$ are not connected continuously by changing $\theta$ from $0$ to $\pi$. 
Since there is no level crossing between $0$ and $\pi$ in this example, this condition suggests that the ground state at $\theta=\pi$ is two-fold degenerate and the time reversal symmetry is spontaneously broken, and this is realized in Fig.~\ref{fig:Z3}.

\subsection{Gauging $\mathbb{Z}_n$ symmetry, 't Hooft anomaly, and global inconsistency}
\label{sec:Zn_gauge}

In order to make the connection between the general discussion in Sec.~\ref{sec:global_inconsistency} and the computation in Sec.~\ref{sec:model1_symmetry}, we rewrite everything using the path integral formalism of this model. 
We discuss the 't Hooft anomaly and global inconsistency of the quantum mechanics~(\ref{eq:model}) in this subsection, and the connection between them will be established in the next subsection. 

To analyze the mixed anomaly or global inconsistency, we promote the global $\mathbb{Z}_n$ symmetry of (\ref{eq:model}) to the local gauge symmetry, and it can be done by coupling the theory (\ref{eq:model}) to a $\mathbb{Z}_n$ topological gauge theory~\cite{Kapustin:2014gua}. 
First, let us write down the continuum description of the $\mathbb{Z}_n$ topological gauge theory, 
\bea
\label{eq:top_action}
S_{\text{top},k}=\im\int F\wedge (\diff B-nA)+\im {k}\int A. 
\eea
Here, $A=A_0\diff \tau$ is the $U(1)$ one-form gauge field, $B$ is the $U(1)$ zero-form gauge field, and $F$ is the zero-form auxiliary field introduced as the Lagrange multiplier. The second term is the one-dimensional Chern--Simons term, and the level $k$ must be an integer for invariance under the $U(1)$ gauge transformation, 
\be
\label{eq:U1_gauge_1}
A\mapsto A+\diff \lambda, \; B\mapsto B+n\lambda, \; F\mapsto F. 
\ee
The level $k$ is identified with $k+n$ because the equation of motion of $F$ gives $n A=\diff B$ and thus $n\int A=\int \diff B\in 2\pi\mathbb{Z}$. We can regard this pair $(A,B)$ as the $\mathbb{Z}_n$ gauge field. 
In order to make the following discussion simpler, we integrate out $F$: The topological action becomes 
\be
\label{eq:topological_Zn_theory}
S_{\mathrm{top},k}[A,B]=\im k \int A, 
\ee
and $B$ dependence appears implicitly through the constraints $n A=\diff B$. 
Next, we couple (\ref{eq:model}) to the topological $\mathbb{Z}_n$-gauge theory (\ref{eq:topological_Zn_theory}) by postulating the following transformation of $q$ under the $U(1)$ guage transformation (\ref{eq:U1_gauge_1}), 
\be
\label{eq:U1_gauge_2}
 q\mapsto q-\lambda.
\ee
The gauge-invariant combinations are $\diff q+A$ and $nq+B$, and thus the gauge-invariant action becomes 
\bea
 \label{eq:gauged_action}
S[q,A,B]&=&\int \diff \tau\left[{1\over 2}(\dot{q}+A_0)^2+V(nq+B)\right]
-{\im\theta\over 2\pi}\int(\diff q+A)+S_{\mathrm{top},k}.  
\eea
We can readily get the partition function $\calZ_{\theta,k}[(A,B)]$ under the background $\mathbb{Z}_n$ gauge field $(A,B)$ as 
\be
\calZ_{\theta,k}[(A,B)]=\int \Diff q \exp(-S[q,A,B]), 
\ee
and the set of couplings is extended by the Chern--Simons level $k\in \mathbb{Z}_n$. 

The time reversal operation $\sfT$ of the Euclidean path integral is chosen as follows:
\be
q(\tau)\mapsto q(-\tau),\; A_0(\tau)\mapsto -A_0(-\tau),\;  B(\tau)\mapsto B(-\tau). 
\ee
The transformation of the dynamical variable $q$ is same as the original one (\ref{eq:time_reversal_q}), and the transformation of background fields are chosen in such a way that the equation of motion is unchanged. 
That is, the covariant derivative $(\dot{q}+A_0)$ is changed to $-(\dot{q}+A_0)$, and $n A=\diff B$ is unchanged under this time reversal transformation. 
Under this transformation, let us check the property of the partition function under the background gauge field at $\theta=0,\pi$. 

The original theory is time reversal invariant at $\theta=0$ and $\pi$. 
At $\theta=0$, the topological $\theta$ term is absent, and thus the $\sfT$ transformation only flips the sign of the Chern-Simons term:
\be
\im k\int A\mapsto -\im k\int A=\im k\int A-2\im k\int A. 
\ee
Therefore, the transformation law of the partition function at $\theta=0$ is 
\be
\label{eq:Z_transformation_0}
\calZ_{0,k}[\sfT\cdot(A,B)]=\calZ_{0,k}[(A,B)]\exp\left(2\im k\int A\right). 
\ee
We can eliminate the additional phase of (\ref{eq:Z_transformation_0}) by choosing appropriate $k$, i.e., 
\be
2k=0 \ \ \ (\mathrm{mod} \ n).
\ee 
When $n$ is even, we have two solutions, $k=0,n/2$ (mod $n$), and when $n(\ge 3)$ is odd, we have the unique solution, $k=0$ (mod $n$). It should be noted that these values of $k$ are identical with the $\mathbb{Z}_n$ charges for singlet states at $\theta=0$ that are calculated in Sec.~\ref{sec:model1_symmetry}. 

At $\theta=\pi$, a nontrivial thing happens because the topological $\theta$ term also flips its sign under time reversal $\sfT$. To see it, let us apply the $\sfT$ transformation to the $\theta$ term at $\theta=\pi$:
\bea
-{\im \pi\over 2\pi}\int (\diff q+A)&\mapsto&{\im \pi\over 2\pi}\int (\diff q+A)\nonumber\\
&=&-{\im \pi\over 2\pi}\int (\diff q+A)+\im \int \diff q+\im \int A. 
\eea
Two additional terms appear after the $\sfT$ transformation of the $\theta$ term; $\int \diff q$ and $\int A$. $\int \diff q$ does not play any role in the path integral, because $\im \int \diff q\in 2\pi \im \mathbb{Z}$. 
Additional $\int A$ shifts the Chern-Simons level by $1$. Combined with the flip of the Chern-Simons term, the $\sfT$ transformation of the partition function at $\theta=\pi$ is obtained as 
\be
\label{eq:Z_transformation_pi}
\calZ_{\pi,k}[\sfT\cdot (A,B)]=\calZ_{\pi,k}[(A,B)]\exp\left(\im (2k-1)\int A\right). 
\ee
In order to preserve the time reversal symmetry under background the $\mathbb{Z}_n$-gauge field, we must choose $k$, such that  
\be
2k-1=0, \ \ \ (\mathrm{mod} \  n). 
\ee
For even $n$, the condition has no solution. The phase factor of (\ref{eq:Z_transformation_pi}) cannot be eliminated by local counter terms, and thus there is the mixed 't Hooft anomaly between $\mathbb{Z}_n$ and the time reversal symmetry. The anomaly matching claims that the ground state must be degenerate at $\theta=\pi$ when $n$ is even. 
For odd $n(\ge 3)$, this has the solution $k=(n+1)/2$ modulo $n$, and no 't Hooft anomaly exists. It should again be noticed that this is same with the $\mathbb{Z}_n$ charge of the singlet state at $\theta=\pi$ as computed in Sec.~\ref{sec:model1_symmetry}. 

For odd $n\ge 3$, there is a global inconsistency between $\theta=0$ and $\theta=\pi$. To eliminate phases at $\theta=0$ and $\theta=\pi$, the Chern-Simons level $k$ should be chosen as 
\be
k_0=0, \ \ k_{\pi}={n+1\over 2}, 
\ee
respectively. We cannot choose simultaneous $k$ eliminating phases because $k_0\not = k_{\pi}$ and $k$ is the discrete parameter. To circumvent it, we need the $2$-dimensional bulk $\Sigma$ with $\p \Sigma=S^1_{\beta}$ as in the case of the anomaly inflow, and then the bulk topological field theory, 
\be
S_{2\mathrm{d},\Sigma}[A]=\im \theta{n+1\over 2\pi}\int_{\Sigma}{\diff A}, 
\ee
can simultaneously eliminate the phases at $\theta=0,\pi$~\cite{Komargodski:2017dmc}. At $\theta=0,\pi$, this topological action is independent of the choice of $\Sigma$ unlike the case of 't~Hooft anomaly, but it is not true for generic $0<\theta<\pi$ and the information of the bulk $\Sigma$ is necessary in order to connect $\theta=0,\pi$. 

\subsection{Relation between two formalisms}

The central extension in operator formalism and local counter terms resulted from gauging global symmetry in path integral formalism seemingly give same information about mixed anomaly and global inconsistency. Here, we show the connection by an explicit computation. We start with the path integral formalism (one can go the other way around) with the action \eqref{eq:gauged_action}. We fix a gauge by requiring $B=0\ (\mathrm{mod}\ 2\pi )$, and the equation of motion $\diff B=nA$ is solved by
\bea
 B=\sum_{i}2\pi \ell_i\Theta(\tau-\tau_i), \ \  A=\sum_{i}\frac{2\pi \ell_i}{n}\delta(\tau-\tau_i)\diff \tau,
\eea
where $\Theta(\tau)$ is the step function and $\delta(\tau)$ is the delta function, for ${\tau_i}\in\mathbb{R}$ and $\ell_i\in\mathbb{Z}$.
Let us calculate the partition function under this background $\mathbb{Z}_n$ gauge field,
\bea
\label{eq:path_integral_operator_correspondence}
\calZ_{\theta,k}[(A,B)] 
 &=&\int\mathcal{D}q\mathcal{D}p
 \exp\Bigg[\int \diff \tau\left(\im p(\dot{q}+A_0)-{1\over 2}\left(p-\frac{\theta}{2}\right)^2-V(nq+B)-\im k A_0\right)\Bigg]\nonumber\\
 &=&\int\mathcal{D}q\mathcal{D}p
 \exp\left[\int \diff \tau\Bigl(\im p\dot{q}-H(p,q)\Bigr) \right]
 \exp\left[\sum_i\frac{2\pi\im \ell_i}{n}(p(\tau_i)-k) \right]
 \nonumber\\
 &=&\left<\prod_i\left(\mathrm{e}^{-2\pi\im{k/n}}\sfU(\tau_i)\right)^{\ell_i}\right>. 
\eea
It can now be explicitly shown the relation between the commutation relation (\ref{eq:com_rel_1}) and the phases in (\ref{eq:Z_transformation_0})  and (\ref{eq:Z_transformation_pi}). Using the commutation relation, we get 
\bea
\sfT\left(\mathrm{e}^{-2\pi\im{k/n}}\sfU\right)\sfT^{-1}=\left\{
 \begin{array}{ll}
  \mathrm{e}^{2\pi \im (2k)/ n}\left(\mathrm{e}^{-2\pi\im{k/n}}\sfU\right), &(\theta=0),
  \\
  \mathrm{e}^{{2\pi\im(2k-1)/n}}\left(\mathrm{e}^{-2\pi\im{k/n}}\sfU\right), \ \  &(\theta=\pi).
 \end{array}
 \right.
\eea
The $\sfT$ transformation acting on the right hand side of (\ref{eq:path_integral_operator_correspondence}) gives the correct additional phases: At $\theta=0$, we get 
\be
\prod_{\im} \left(\mathrm{e}^{2\pi \im (2k)/n}\right)^{\ell_i}=\exp\left(2\im k\int A\right), 
\ee
and, at $\theta=\pi$, we get 
\be
\prod_{\im} \left(\mathrm{e}^{2\pi \im (2k-1)/n}\right)^{\ell_i}=\exp\left(\im (2k-1)\int A\right).  
\ee
We should emphasize that the phase factors which come from the local counter term are precisely same as those appear as a central extension. 


\section{Quantum mechanics of two particles on $S^1$}\label{sec:model2}

We consider the quantum mechanics with the target space $U(1)\times U(1)$ corresponding to two distinguishable particles moving on a ring with flux threading. 
We shall go through the parallel argument as we have done in the last section, but this model exhibits new ingredients and the global inconsistency plays a particularly important role. 
The Euclidean classical action is 
\be
\label{eq:model2}
S[q_1,q_2]=\int\diff \tau \left[{1\over 2}(m_1\dot{q}_1^2+m_2\dot{q}_2^2)+V(q_1-q_2)\right]-{\im\theta_1\over 2\pi}\int\diff q_1-{\im\theta_2\over 2\pi}\int\diff q_2,
\ee
where $m_1$ and $m_2$ are distinct mass parameters for each particle and the potential $V(x)$ is represented as the Fourier series \eqref{eq:potential1}, which is a smooth $2\pi$ periodic function.
Each $q_i\ (i=1,2)$ is the map $q_i:S^1_{\beta}\to \mathbb{R}/2\pi\mathbb{Z}$. 
The theta parameters $\theta_i$ are $2\pi$ periodic variables.

With use of the path integral the partition function is expressed as
\be
 \label{eq:path_integral2}
 \calZ_{(\theta_1,\theta_2)}=\int\Diff q_1\Diff q_2\exp(-S[q_1,q_2]).
\ee

In the operator formalism, the partition function is expressed as $\calZ_{(\theta_1,\theta_2)}=\text{tr}_\mathcal{H}[\exp(-\beta\hat{H})]$ with the hamiltonian given by
\be
 \hat{H}(\hat{p}_1,\hat{q}_1,\hat{p}_2,\hat{q}_2)
 =\frac{1}{2m_1}\left(\hat{p}_1-\frac{\theta_1}{2\pi}\right)^2+\frac{1}{2m_2}\left(\hat{p}_2-\frac{\theta_2}{2\pi}\right)^2+V(\hat{q}_1-\hat{q}_2).
\ee
where $[\hat{q}_i,\hat{p}_j]=\im \delta_{ij}$.

\subsection{Symmetries, central extension, and global inconsistency}

The action \eqref{eq:model2} possesses $U(1)$ symmetry generated by
\be
 \sfU_\alpha: q_i(\tau) \mapsto q_i(\tau)+\alpha,
\ee
where $i=1,2$ and $\alpha$ is a $2\pi$ periodic constant, i.e., $\sfU_{2\pi}=1$. The corresponding generator is given by $\sfU_\alpha=\mathrm{e}^{\im \alpha(\hat{p}_1+\hat{p}_2)}$ and satisfies a commutation relation $\sfU_{\alpha}\hat{H}\sfU_{-\alpha}=\hat{H}$.
The time reversal transformation
\be
 \sfT: q_i(\tau)\mapsto q_i(-\tau), \ \dot{q}_i(\tau)\mapsto -\dot{q}_i(-\tau),
\ee
becomes an additional symmetry at $(\theta_1,\theta_2)=(0,0), (0,\pi), (\pi,0), (\pi,\pi)$, which are the high symmetry points of the model.

We analyze commutation relations of $\sfU_\alpha$ and $\sfT$ to study the 't~Hooft anomaly and global inconsistency.
Let the high symmetry points be denoted by $(\theta_1,\theta_2)=(j_1\pi,j_2\pi)$ with $j_1,j_2\in\mathbb{Z}$. The condition for the time-reversal symmetry, $\sfT\hat{H}\sfT^{-1}=\hat{H}$, combined with anti-unitarity $\sfT\im\sfT=-\im$ requires $\sfT \hat{q}_i \sfT^{-1}=\hat{q}_i\, (i=1,2)$ and
\be
\sfT \hat{p}_1\sfT^{-1}=-\hat{p}_1+j_1, \ \ \sfT \hat{p}_2\sfT^{-1}=-\hat{p}_2+j_2, \ \ (\theta_1,\theta_2)=(j_1\pi,j_2\pi).
\ee
%
Therefore, the commutation relations between $\sfU$ and $\sfT$ are,
\be
\label{eq:com_rel}
\sfT\sfU_\alpha\sfT^{-1}=
  \mathrm{e}^{\im(j_1+j_2)\alpha}\sfU_\alpha, \ \ (\theta_1,\theta_2)=(j_1\pi,j_2\pi). 
\ee
At $(\theta_1,\theta_2)=(0,0)$, we obtained expected relation from $U(1)\times\mathbb{Z}_2$ symmetry.
At $(\theta_1,\theta_2)=(0,\pi)$ and $(\pi,0)$, we have additional phase factor $\mathrm{e}^{\im \alpha}$. We again try to absorb it by redefining the operator $\sfU'_\alpha\equiv\mathrm{e}^{-\im\alpha/2}\sfU_\alpha$. But $\sfU'_\alpha$ forces the periodicity of $\alpha$ to be extended to $4\pi$. Thus, avoiding the central extension necessarily yields the double cover of $U(1)\times\mathbb{Z}_2$ and this is a symptom of a mixed 't~Hooft anomaly. Although similar issue seems to appear at $(\theta_1,\theta_2)=(\pi,\pi)$ this is not true because the phase factor $\mathrm{e}^{2\im\alpha}$ can be absorbed by a redefinition $\sfU''_{\alpha}\equiv\mathrm{e}^{-\im\alpha}\hat{\mathcal{O}}_{\alpha}$ without extending periodicity of $\alpha$.
It is again noted that, although there is no central extension at $(\theta_1,\theta_2)=(0,0)$ and $(\pi,\pi)$ respectively, we cannot choose common operator $\sfU_\alpha$ (or $\sfU''_\alpha$). This is the global inconsistency.
So far, we found similar observations as those we saw in the last section.

An interesting thing happens at $(\theta_1,\theta_2)=(\pi,-\pi)$; $\sfT \hat{p}_1 \sfT^{-1}=-\hat{p}_1+1$ and $\sfT \hat{p}_2 \sfT^{-1}=-\hat{p}_2-1$ lead to a commutation relation
\be
\label{eq:com_rel2}
\sfT \sfU_\alpha\sfT^{-1}=\sfU_\alpha.
\ee
Hence, there is not a mixed 't~Hooft anomaly and also a global inconsistency does not exist between $(0,0)$ and $(\pi,-\pi)$. A global inconsistency however exists between $(0,0)$ and $(\pi,\pi)$. It is noted that the theory at $(\pi,-\pi)$ must show the same property as one at $(\theta_1,\theta_2)=(\pi,\pi)$ because $\theta_2$ is $2\pi$ periodic parameter.\footnote{Of course the same is true at $(\theta_1,\theta_2)=(-\pi,\pi)$ by using $2\pi$ periodicity of $\theta_1$.}
This observation is not a contradiction and yields an important constraint on the energy spectrum as we will see momentarily.

\begin{figure}[t]
\centering
\includegraphics[scale=0.7]{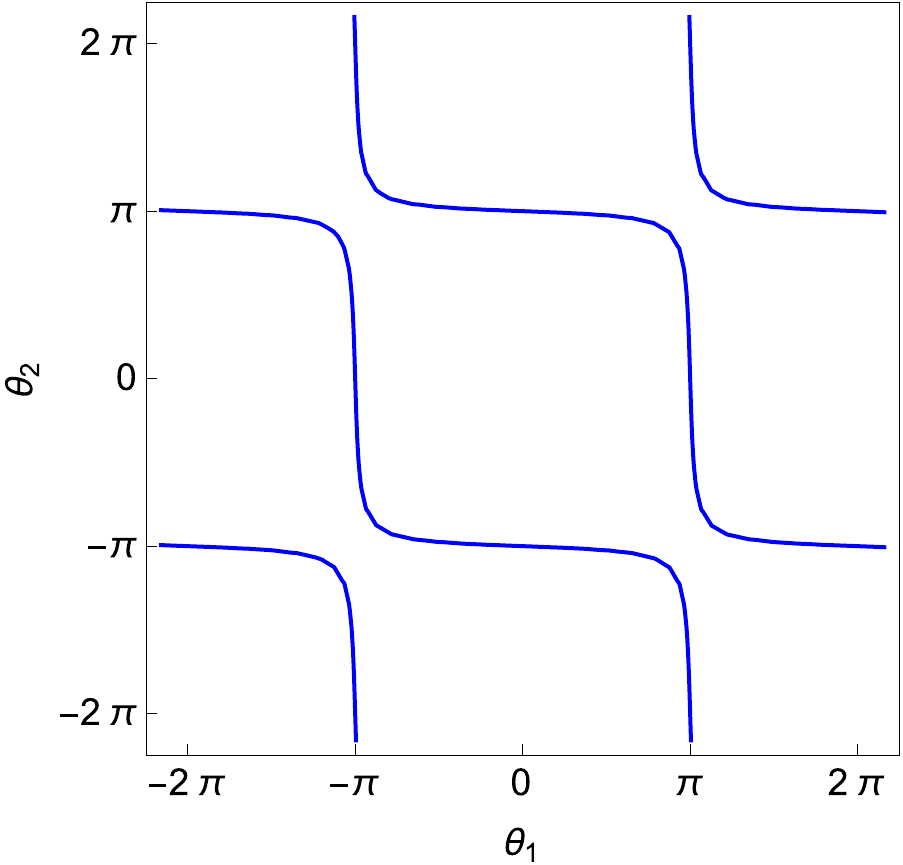}
\caption{Phase diagram on $(\theta_1,\theta_2)$-plane with $\lambda=0.2$. Each line represents a level crossing (phase transition). The phase structure is $2\pi$ periodic along $\theta_1$ and $\theta_2$ axises.}
\label{fig:phase}
\end{figure}

We explore the implication of the above argument to energy spectrum and phase diagram. The same argument as we gave in the last section results in the existence of degenerate state at $(\theta_1,\theta_2)=(0,\pi),(\pi,0)$ and we do not repeat here. Instead, we restrict our attention to $(\theta_1,\theta_2)=(0,0),(\pi,\pi),(\pi,-\pi)$.
The simultaneous eigenstate of $\sfU_\alpha$ and $\sfT$ would satisfy
\be
 \sfU_\alpha|\psi\rangle=\mathrm{e}^{\im\alpha k}|\psi\rangle, \  \sfT|\psi\rangle=\eta|\psi\rangle,
\ee
where $k\in\mathbb{Z}$ because $\sfU_{2\pi}=1$.
Then, by using the commutation relations \eqref{eq:com_rel} and \eqref{eq:com_rel2}, the parallel discussion given in Sec.~\ref{sec:model1_symmetry} leads to the following $U(1)$ transformation law of states,
\be
 \sfU_\alpha|\psi_{(0,0)}\rangle=|\psi_{(0,0)}\rangle, \ 
 \sfU_\alpha|\psi_{(\pi,\pi)}\rangle=\mathrm{e}^{i\alpha}|\psi_{(\pi,\pi)}\rangle, \ 
 \sfU_\alpha|\psi_{(\pi,-\pi)}\rangle=|\psi_{(\pi,-\pi)}\rangle,
\ee
at $(\theta_1,\theta_2)=(0,0),(\pi,\pi),(\pi,-\pi)$, respectively.
Since $|\psi_{(\pi,\pi)}\rangle$ has different $U(1)$ charge from $|\psi_{(0,0)}\rangle$ and $|\psi_{(\pi,-\pi)}\rangle$, $|\psi_{(\pi,\pi)}\rangle$ cannot be continuously connected to the other two states at high symmetry points.
In addition, $(\pi,\pi)$ and $(\pi,-\pi)$ must be identified because $\theta_2$ is a $2\pi$ periodic parameter as we mentioned before.
The compatible consequence is that $(\pi,\pi)$ and $(\pi,-\pi)$ are separated by a phase transition as shown in Fig.~\ref{fig:phase}.
Otherwise, $\sfT$-invariant state at $(\pi,\pi)$ would be connected to $\sfT$-broken state at $(\pi,-\pi)$ without a level crossing, which contradicts to the fact that $(\pi,\pi)$ and $(\pi,-\pi)$ must have identical energy spectra.

The above arguments are indeed checked by a explicit computation of the energy spectra with a specific potential
\be
 V(q_1-q_2)=\lambda\cos(q_1-q_2).
\ee

%
%
%

\begin{figure}[t]
\centering
\begin{minipage}{.49\textwidth}
\subfloat[$\theta_2=0$.]{
\includegraphics[scale=0.2]{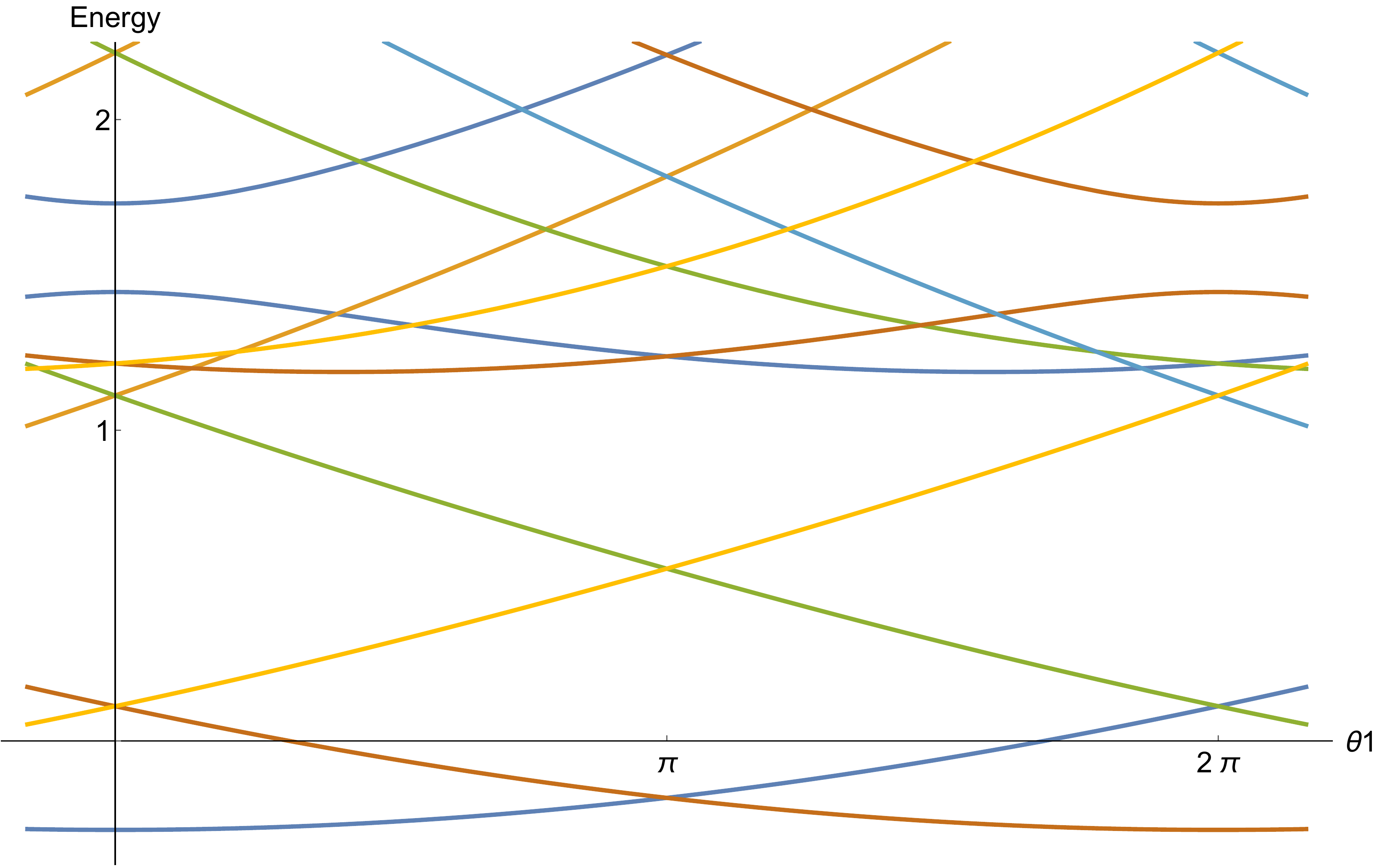}
\label{fig:energy_qm_bifund_theta1}
}
\end{minipage}\
\begin{minipage}{.49\textwidth}
\subfloat[$\theta=\theta_1=\theta_2$.]{
\includegraphics[scale=0.2]{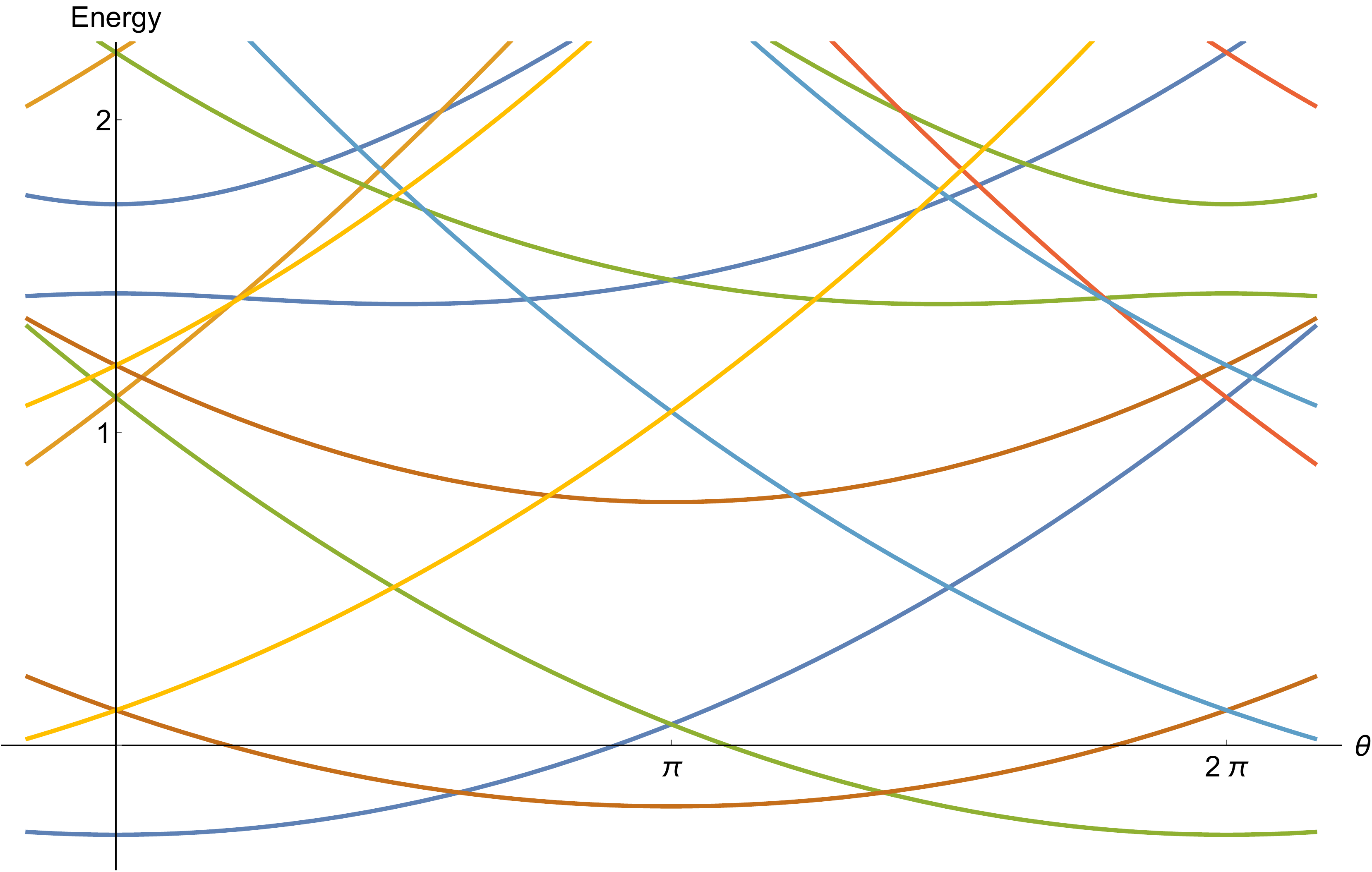}
\label{fig:energy_qm_bifund_theta}
}
\end{minipage}
\begin{minipage}{.49\textwidth}
\subfloat[$\theta'=\theta_1=-\theta_2$.]{
\includegraphics[scale=0.2]{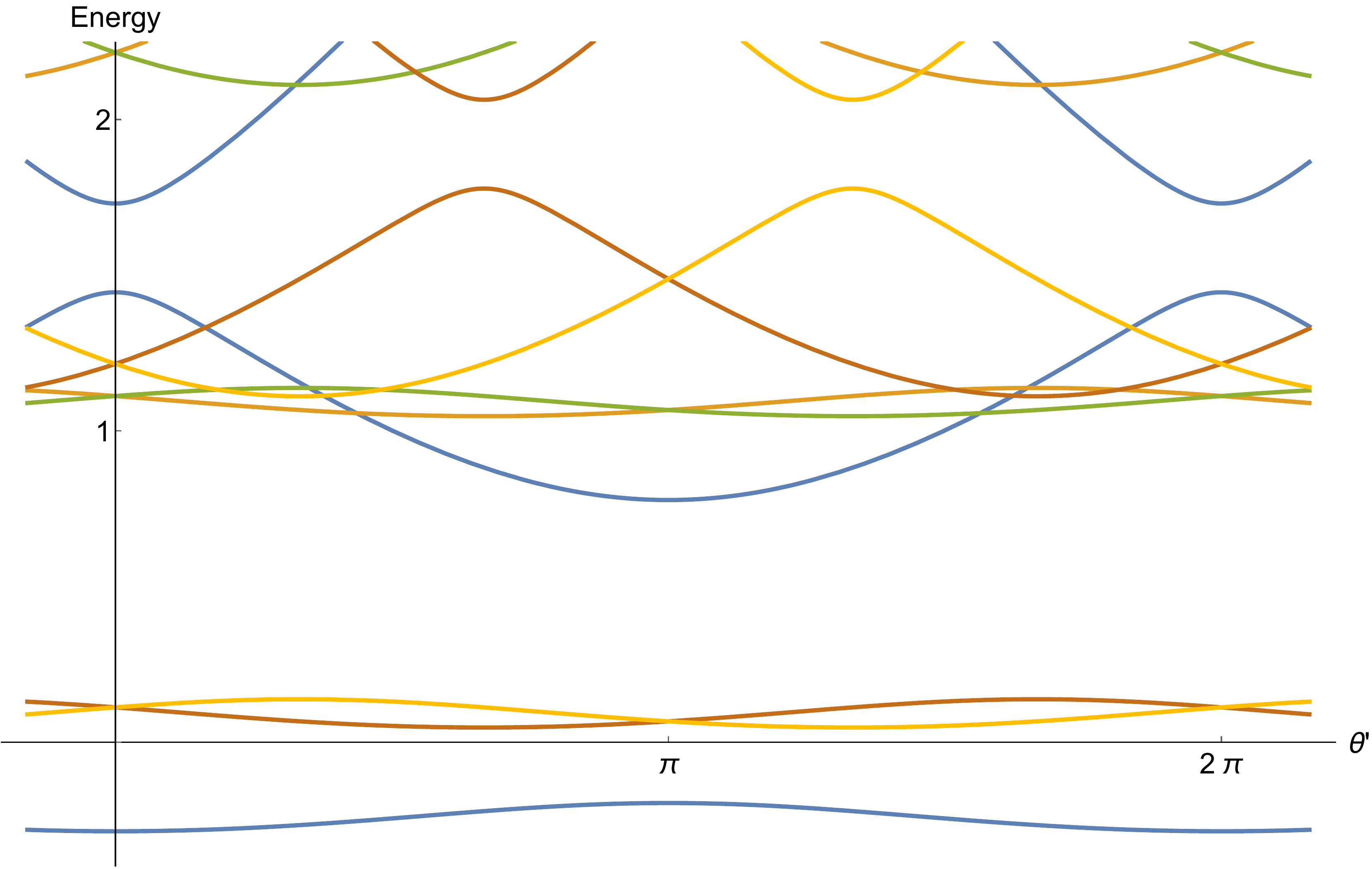}
\label{fig:energy_qm_bifund_theta'}
}
\end{minipage}
\caption{Energy spectra as functions of $\theta$ with $m_1=1$, $m_2=1/2$ and $\lambda=1$. Color of lines indicates the $U(1)$ charge of states. 
(a) All the levels are degenerate at $(\theta_1,\theta_2)=(\pi,0)$ due to the 't~Hooft anomaly.
(b) A singlet state at $(\theta_1,\theta_2)=(0,0)$ must be connected to a degenerate state at $(\theta_1,\theta_2)=(\pi,\pi)$ and vice versa due to the global inconsistency.
(c) Singlet states at $(\theta_1,\theta_2)=(0,0)$ are connected to singlet states at $(\theta_1,\theta_2)=(\pi,-\pi)$.}
\label{fig:energy_qm_bifund}
\end{figure}

As shown in Fig.~\ref{fig:energy_qm_bifund_theta1}, all the states at $(\theta_1,\theta_2)=(\pi,0)$ form pairs and the time reversal symmetry is spontaneously broken.

Fig.~\ref{fig:energy_qm_bifund_theta} shows the energy spectra as function of $\theta_1=\theta_2=\theta$. If a nondegenerate state exists at $\theta=0$, it is continuously connected to a degenerate state at $\theta=0$ (see the lowest blue curve in Fig.~\ref{fig:energy_qm_bifund_theta}, for instance) and vice versa (the lowest brown curve).
Interestingly, the vacuum (lowest-energy) states are singlet both at $\theta=0$ and $\theta=\pi$, which is allowed because the level crossing (phase transition) separates these high symmetry points.
The $U(1)$ charge of the lowest-energy state can jump at the crossing point since the points are not continuously connected by changing $\theta$.
This is the new ingredient which we did not see in the last section. Namely, the global inconsistency does not necessarily lead to the existence of degenerate vacuum at high symmetry points.
Therefore, This result does not contradict to the fact that there is a global inconsistency between $(\theta_1,\theta_2)=(0,0)$ and $(\pi,\pi)$.

Finally, we see that the vacuum is nondegenerate for $\theta_1=-\theta_2=\theta'$ (Fig.~\ref{fig:energy_qm_bifund_theta'}),
which is consistent with the discussion in the last section that there is neither an 't~Hooft anomaly nor global inconsistency at $\theta_1=-\theta_2=\pi$.

A phase diagram on $(\theta_1,\theta_2)$-plane (Fig.~\ref{fig:phase}) follows the energy spectrum and level crossing computed above.
As expected from the 't~Hooft anomaly, Level crossing lines pass at $(0,\pm\pi)$ and $(\pm\pi,0)$.
High symmetry points $(\theta_1,\theta_2)=(0,0)$ and $(\pi,-\pi)$ are connected without level crossing while $(0,0)$ and $(\pi,\pi)$ are separated by a level crossing line, which agrees with our consideration based on the global inconsistency.

Based on the constraints from the global inconsistency, we could come up with a little more exotic phase diagram which we did not find here. The other possibility we could draw from the global inconsistency between $(0,0)$ and $(\pi,\pi)$ is that the nondegenerate vacuum at $(0,0)$ is connected to the degenerate vacua at $(\pi,\pi)$ without level crossing. Then, there exists the degenerate vacua at $(\pi,-\pi)$ as well due to the $2\pi$ periodicity of $\theta_2$. Therefore,  the points $(0,0)$ and $(\pi,-\pi)$ must be separated by another level crossing line because the singlet state at $(0,0)$ cannot be connected to the $\sfT$-broken state at $(\pi,-\pi)$ due to the absence of global inconsistency between these two points. See Ref.~\cite{Tanizaki:2017bam} for detailed discussion on $SU(n)\times SU(n)$ bifundamental gauge theory with two $\theta$ parameters corresponding to two gauge groups.
In the theory, the almost same conditions are obtained by using global inconsistency and 't~Hooft anomaly and two possible diagrams are proposed,
and our phase diagram Fig.~\ref{fig:phase} actually fits to one of the proposal made in \cite{Tanizaki:2017bam}.

\subsection{Gauging $U(1)$ symmetry, 't~Hooft anomaly, and global inconsistency}

We promote the global $U(1)$ symmetry to the local gauge symmetry by coupling to the background $U(1)$ gauge field $A$ in order to study the 't~Hooft anomaly and global inconsistency for $U(1)\times\mathbb{Z}_2$ symmetry. To this end, we study the model \eqref{eq:model2} in the path integral formalism \eqref{eq:path_integral2} as we have done in Sec.~\ref{sec:Zn_gauge}. 
The topological $U(1)$ gauge theory we need to couple here is
\be
 S_{\mathrm{top},k}[A]=\im k\int A,
\ee
which is $U(1)$ level-$k$ Chern-Simons term in one dimension. The invariance under $U(1)$ gauge transformation
\be
 A\mapsto A+\diff\lambda
\ee
requires the level to be an integer, $k\in\mathbb{Z}$.
By postulating $U(1)$ gauge transformation, 
\be
q_1\mapsto q_1+\lambda, \ q_2\mapsto q_2+\lambda, 
\ee
we obtain the gauge invariant action coupled to the topological gauge theory,
\bea
S[q_1,q_2,A]&=&\int \diff \tau\left[\frac{m_1}{2}(\dot{q}_1+A_0)^2+\frac{m_2}{2}(\dot{q}_2+A_0)^2+V(q_1-q_2)\right]\nonumber\\
&&-{\im\theta_1\over 2\pi}\int (\diff q_1+A)-{\im\theta_2\over 2\pi}\int (\diff q_2+A)+S_{\mathrm{top},k}[A]. 
\eea
Therefore, the partition function coupled to the background $U(1)$ gauge field is given by
\be
 \calZ_{(\theta_1,\theta_2),k}[A]=\int\mathcal{D}q_1\mathcal{D}q_2\exp{(-S[q_1,q_2,A])}
\ee
We will see how the partition function at high symmetry points transforms under time reversal operation,
\be
 q_1(\tau)\mapsto q_1(-\tau), \ q_2(\tau)\mapsto q_2(-\tau), \ A_0(\tau)\mapsto -A_0(-\tau).
\ee

At $(\theta_1,\theta_2)=(0,0)$, the partition function transforms as
\be
\calZ_{(0,0),k}[\sfT\cdot A]=\calZ_{(0,0),k}[A]\exp\left(2\im k\int A\right).
\ee
The time reversal invariance requires $k=0$. Notice that the same transformation law holds at $(\theta_1,\theta_2)=(\pi,-\pi)$, which also results in $k=0$.

As we saw in the last section, the transformation of the partition function at $(\theta_1,\theta_2)=(0,\pi)$
\be
\calZ_{(\pi,0),k}[\sfT\cdot A]=\calZ_{(\pi,0),k}[A]\exp\left(\im(2k-1)\int A\right),
\ee
leads to a nontrivial consequence. The time reversal invariance requires $2k-1=0$. Since this condition cannot be satisfied with integer $k$, the time reversal invariance cannot be preserved after gauging the $U(1)$ symmetry. Hence, an 't~Hooft anomaly exists at $(\theta_1,\theta_2)=(0,\pi)$. Clearly, the same is true at $(\theta_1,\theta_2)=(\pi,0)$.

Finally, at $(\theta_1,\theta_2)=(\pi,\pi)$, the partition function transforms as
\be
\calZ_{(\pi,\pi),k}[\sfT\cdot A]=\calZ_{(\pi,\pi),k}[A]\exp\left(\im(2k-2)\int A\right).
\ee
In this case, the time reversal invariance is unbroken by choosing $k=1$, meaning that there exists no mixed 't~Hooft anomaly.
By observing the resulting Chern-Simons levels at $(0,0),(\pi,-\pi),(\pi,\pi)$
\be
 k_{(0,0)}=0=k_{(\pi,-\pi)}, \ k_{(\pi,\pi)}=1,
\ee
we conclude that there are global inconsistencies between $(0,0)$ and $(\pi,\pi)$, and between $(\pi,-\pi)$ and $(\pi,\pi)$, respectively.

It is impossible to eliminate the phases coming out of 't~Hooft anomalies and global inconsistencies by the local counter term, and we need the $2$-dimensional bulk $\Sigma$ with $\p \Sigma=S^1_{\beta}$ to do it keeping the gauge invariance. The $2$-dimensional topological action, 
\be
S_{2\mathrm{d},\Sigma}[A]=\im{(\theta_1+\theta_2)\over 2\pi}\int_{\Sigma} \diff A, 
\ee
cancel additional phases of the partition function. At $(\theta_1,\theta_2)=(\pi,0),(0,\pi)$, this topological action depends on the topology of $\Sigma$, and this detects the mixed 't Hooft anomaly. At $(\theta_1,\theta_2)=(\pi,\pi)$, this does not depend on the choice of $\Sigma$, but the information of the bulk is necessary to connect it with $(\theta_1,\theta_2)=(0,0)$, and this is the signal for the global inconsistency. 

\subsection{More on $\mathbb{Z}_n\times\mathbb{Z}_2$ mixed anomaly}

Finally, we briefly look at the model \eqref{eq:model2} with one-particle potentials $V(nq_1)$ and $V(n q_2)$ in addition to the inter-particle potential $V(q_1-q_2)$, which are represented as Fourier series \eqref{eq:potential1} with different sets of parameters. $V(nq_1)$ and $V(n q_2)$ explicitly break $U(1)$ symmetry down to $\mathbb{Z}_n$ symmetry, which is generated by
\be
\sfU: q_1\mapsto q_1+\frac{2\pi}{n}, \ q_2\mapsto q_2(\tau)+\frac{2\pi}{n}.
\ee
The potentials $V(nq_1)$, $V(n q_2)$ change the conditions for the 't~Hooft anomaly and global inconsistency at high symmetry points.
Here, we do not repeat the operator formalism but present only the path integral formalism. To this end, we promote the global $\mathbb{Z}_n$ symmetry by following the procedure employed in Sec.~\ref{sec:Zn_gauge}.
The topological gauge theory we need to couple is \eqref{eq:topological_Zn_theory} by introducing $\mathbb{Z}_n$ one-form $A$ and $U(1)$ zero-form gauge fields $B$ with the constraint $n A=\diff B$. The $\mathbb{Z}_n$ gauge transformation is given by \eqref{eq:U1_gauge_1} and 
\be
 q_1\mapsto q_1-\lambda, \ q_2\mapsto q_2-\lambda.
\ee
The action invariant under the gauge transformation takes the following form,
\bea
\label{eq:zp_action}
&&S_{(\theta_1,\theta_2),k}[q_1,q_2,A,B]
\nonumber\\
&&=\int \diff \tau\left[\frac{m_1}{2}(\dot{q}_1+A_0)^2+\frac{m_2}{2}(\dot{q}_2+A_0)^2+V(q_1-q_2)+V(nq_1+B)+V(nq_2+B)\right]\nonumber\\
&&-{\im\theta_1\over 2\pi}\int(\diff q_1+A)-{\im\theta_2\over 2\pi}\int\theta_2(\diff q_2+A) +S_{\mathrm{top},k}[A],
\eea

Here, we list the condition for the discrete parameter $k$ at each high symmetry points required by invariance under the time reversal symmetry:
\bea
\label{eq:list}
\left\{
\begin{array}{ll}
k=-k,&(\theta_1,\theta_2)=(0,0),(\pi,-\pi)\\
k=-k+1,&(\theta_1,\theta_2)=(\pi,0),(0,\pi),\\
k=-k+2,&(\theta_1,\theta_2)=(\pi,\pi).
\end{array}
\right. \ \ \ \ (\mathrm{mod}\ n)
\eea
These restrictions result in the following consequences: 
For odd $n\ge3$, an 't~Hooft anomaly does not exist at any high symmetry point. In this case, global inconsistencies exists among $(0,0)$, $(\pi,0),(0,\pi)$ and $(\pi,\pi)$ because
\be
 k_{(0,0)}=0,\ k_{(\pi,0)}=\frac{n+1}{2}=k_{(0,\pi)}, \ k_{(\pi,\pi)}=1,
\ee
which take different values.

For even $n\ge4$, 't~Hooft anomalies appear at $(\theta_1,\theta_2)=(\pi,0),(0,\pi)$ because there is no integer solution for $k$, i.e., the gauge invariance cannot be maintained. Although there is no mixed anomaly at $(\theta_1,\theta_2)=(0,0),(\pi,\pi),(\pi,-\pi)$, a global inconsistency exists between $(0,0)$ and $(\pi,\pi)$ and between $(\pi,-\pi)$ and $(\pi,\pi)$ because
\be
 k_{(0,0)}=0=k_{(\pi,-\pi)}, \ k_{(\pi,\pi)}=1.
\ee

In $n=2$ case, the first and third conditions in \eqref{eq:list} are equivalent mod $n$. Hence, there is no global inconsistency although we still have 't~Hooft anomalies at $(\theta_1,\theta_2)=(\pi,0),(0,\pi)$.

\section{Conclusion}\label{sec:conclusion}

We have clarified the nature of the global inconsistency in comparison with the 't~Hooft anomaly and analyzed their implication on energy spectra by looking at quantum mechanical models.
The 't~Hooft anomaly shows up as an obstruction for gauging a global $G$ symmetry of the system and inevitably leads to nontrivial infrared theories. The global inconsistency has similar nature in that it also appears as an obstruction for gauging symmetry and imposes constraints on the low-energy theory.
The global inconsistency, however, plays a role in more restricted situations, where there exist high symmetry points connected each other by continuous parameters of the theory. The constraints obtained from the global inconsistency is milder than those from the 't~Hooft anomaly due to the fact that it does not necessarily rule out the realization of trivial vacuum at high symmetry points. When there is a global inconsistency between two high symmetry points, one can draw a constraint that the vacuum is nontrivial at either of the points, or that those two points are separated by a phase transition.

We carefully analyzed quantum mechanical models which exhibits 't~Hooft anomalies and global inconsistencies at high symmetry points of the parameter space spanned by theta parameters.
We studied them by the operator formalism and path-integral formalism.
In the operator formalism, by studying central extensions of the symmetry group, one can tell how (non-accidental) level crossings appears in energy spectrum.
In the path-integral formalism, 't~Hooft anomalies and global inconsistencies are detected by gauging a global symmetry as we discussed in Sec.~\ref{sec:global_inconsistency}.
We then established a precise connection between these two formalisms in the quantum mechanical models, which allows us to predict the level crossing in energy spectra by studying 't~Hooft anomalies and global inconsistency. It is noted that the 't~Hooft anomaly matching argument constraints only vacuum property of the QFT because of the assumption on locality of low-energy effective theories. However, they becomes more restrictive in quantum mechanics and one can extract the information on excited states as well by combining the observations drawn from the central extension of symmetry groups.

More specifically, we analyzed the following quantum mechanical models in detail:
In the model describing a particle on a ring, the symmetry group is $\mathbb{Z}_n\times\mathbb{Z}_2$ at high symmetry points $\theta=0$ and $\pi$. 
There is a mixed anomaly at $\theta=\pi$ for even $n$, and a global inconsistency between $\theta=0$ and $\pi$ for odd $n$.
This model is a reminiscent of $SU(n)$ pure Yang-Mills model at $\theta=\pi$ with $\mathbb{Z}_n$ one-form center symmetry and time reversal symmetry.
The second model with two particles is a reminiscent of $SU(n)\times SU(n)$ gauge theory with bifundamental matters. 
There are mixed anomalies at $(\theta_1,\theta_2)=(0,\pi),(\pi,0)$. The global inconsistency appears between $(\theta_1,\theta_2)=(0,0)$ and $(\pi,\pi)$ but not between $(\theta_1,\theta_2)=(0,0)$ and $(\pi,-\pi)$ which indeed agrees with the phase diagram for this model in $(\theta_1,\theta_2)$ plane.
The interesting observation which was absent in the first model is that the global inconsistency does not imply the existence of degenerate vacua at $(\theta_1,\theta_2)=(\pi,\pi)$.
Instead the high symmetry points, $(\theta_1,\theta_2)=(0,0)$ and $(\pi,\pi)$, are separated by a level crossing line in the $(\theta_1,\theta_2)$ space.

In this paper, we observed the consequence of global inconsistency in quantum mechanics, and it would be interesting to see solvable examples of the quantum field theories with global inconsistency. 
As we mentioned in the introduction, the first quantum mechanical model mimics the $SU(n)$ pure Yang--Mills theory~\cite{Gaiotto:2017yup} or the Abelian-Higgs model~\cite{Komargodski:2017dmc,Komargodski:2017smk}, and the second one does the $SU(n)\times SU(n)$ bifundamental gauge theory with massive matter fields~\cite{Tanizaki:2017bam}. 
Adiabatic compactification of these gauge theories with the double-trace deformation or with appropriate matter contents enables us to compute the phase structure based on the controllable semiclassical approximations~\cite{Shifman:2008ja, Unsal:2007vu, Kovtun:2007py, Unsal:2007jx, Unsal:2008ch, Shifman:2009tp, Argyres:2012ka, Argyres:2012vv, Dunne:2012ae, Dunne:2012zk, Poppitz:2012sw, Anber:2013doa, Cherman:2013yfa, Cherman:2014ofa, Misumi:2014bsa, Misumi:2014jua, Dunne:2016nmc, Cherman:2016hcd, Fujimori:2016ljw, Kozcaz:2016wvy, Sulejmanpasic:2016llc, Yamazaki:2017ulc, Aitken:2017ayq}, and thus the resurgence theory on QFT will provide a lot of examples to deepen our understandings on global inconsistency condition.  


\section*{Acknowledgment}
Y.~K. is supported by the Grants-in-Aid for JSPS fellows (Grant No.15J01626). Y.~T. is financially supported by RIKEN special postdoctoral program.





\bibliographystyle{JHEP}
\bibliography{./QFT,./QM}
%

\end{document}